\documentclass[twocolumn,showpacs,preprintnumbers,amsmath,amssymb]{revtex4}
\usepackage{tipa}
\usepackage{graphicx}
\usepackage{dcolumn}
\usepackage{bm}

\begin{document}

\title{Implications to $c\bar s$ assignments of $D_{s1}(2700)^\pm$ and
$D_{sJ}(2860)$}
\author{Ailin Zhang$^{1,2}$\footnote{Email:
zhangal@staff.shu.edu.cn}} \affiliation{$^1$Department of Physics,
Shanghai University, Shanghai 200444, China\\
$^2$Theoretical Physics Center for Science Facilities(TPCSF), CAS,
Beijing 100049, China}

%\date{\today}

\begin{abstract}
Possible assignments of $D_{s1}(2700)^\pm$ and $D_{sJ}(2860)$ in the
conventional quark model are analyzed. The study indicates that both
the orbitally excited $c\bar s$ and the radially excited $c\bar s$
are possible. Some implications to these assignments are explored.
If $D_{s1}(2700)^\pm$ and $D_{sJ}(2860)$ are the orbitally excited
D-wave $1^-~(j^P={3\over 2}^-)$ and $3^-~(j^P={5\over 2}^-)$,
respectively, another orbitally excited D-wave $2^-$ $D_s$,
$D_{s2}(2800)$, is expected. If $D_{s1}(2700)^\pm$ and
$D_{sJ}(2860)$ are the first radially excited $1^-~(j^P={1\over
2}^-)$ and $0^+~(j^P={1\over 2}^+)$, respectively, other two
radially excited $0^-~D^\prime_s(2582)$ and
$1^+~D^\prime_{s1}(2973)$ are expected. $D_{s2}(2800)$ and
$D^\prime_{s1}(2973)$ are mixing states. The chiral doubling
relation may exist in radially excited $D_s$, the splitting between
the parity partners(the $(0^-,1^-)$ and the $(0^+,1^+)$) is $\approx
280$ MeV.
\end{abstract}
\pacs{11.55.Jy, 12.40.Nn, 12.40.Yx\\
Keywords: Charmed strange mesons, Mixing effect, Decay, Regge
trajectory}

\maketitle

\section{Introduction}
Heavy-light($Q\bar q$) mesons lie between the light ($q\bar q$)
mesons and the heavy $Q\bar Q$ mesons. Both the heavy quark symmetry
and the light quark chiral symmetry apply in this energy region.
These symmetries(and symmetry violations) may result in some special
features to the spectrum, decay and production of the heavy-light
mesons. The study of heavy-light mesons is helpful to detect the
properties of quark dynamics.

Heavy-light mesons have been systematically explored in relativized
quark model~\cite{gi,gk}, heavy quark symmetry theory~\cite{hqet},
relativistic quark model~\cite{rq}, chiral quark
model~\cite{pe,zhao}, lattice~\cite{latt}, coupled channels
models~\cite{cc1,cc2} and other
models~\cite{other1,other2,other3,other4,other5}. For most
heavy-light mesons, theoretical predictions of the masses and the
decay characters are comparable with experiments.

In recent years, some "exotic" heavy-light mesons have been
observed. For the "exotic" mesons, the measured masses or the
observed decay properties are very different with theoretical
predictions. So far, no "exotic" charmed meson($c\bar q(q=u,d)$) has
been reported, and only "exotic" charmed strange mesons($c\bar s$)
have been observed~\cite{pdg08}.

$D^{\star}_{s0}(2317)^\pm$ was firstly observed by
BaBar~\cite{babar1} in $D^\star_{s0}(2317)\to D^+_s\pi^0$ with mass
near $2.32~GeV$, $\approx 40$ MeV below $DK$ threshold. It has full
width $\Gamma<3.8$ MeV at $95\%$ confidence level. This state was
subsequently confirmed by CLEO~\cite{cleo1} and BELLE~\cite{belle}.
For the lower mass and the narrower width, controversial
interpretations of this state were proposed.

$D_{s1}(2460)^\pm$ was firstly reported by CLEO~\cite{cleo1} in
$D_{s1}(2460)^\pm\to D^\star_s\pi^0$. It was also observed by
BELLE~\cite{belle} and BaBar~\cite{babar2}. This state has mass
$2459.6\pm 0.6$ MeV, $\approx 50$ MeV below $D^\star K$ threshold,
and full width $\Gamma<3.5$ MeV at $95\%$ CL. Similar to
$D^{\star}_{s0}(2317)^\pm$, the interpretations of
$D_{s1}(2460)^\pm$ are of controversial.

Though $D^{\star}_{s0}(2317)^\pm$ and $D_{s1}(2460)^\pm$ have
"exotic" features, their exotic features could be explained in the
traditional $c\bar s$ quark model. At present, most people believe
that these two states belong to the P-wave multiplets of the $D_s$.
They are thought of as the chiral doubler of $D_s(1969)^\pm$ and
$D^\star_s(2112)^\pm$. It is very possible that the lowest S-wave
and P-wave  multiplets of $D_s$ have been established.

While the P-wave multiplets have been filled, the higher $D_s$ are
the D-wave $c\bar s$ or the radially excited $c\bar s$.

Very recently, two new states with higher energy were observed.
$D_{s1}(2700)^\pm$ was firstly observed by Belle~\cite{belle2} in
$B^+\to \bar D^0D_{s1}\to\bar D^0D^0K^+$ with $M=2715\pm
11^{+11}_{-14}$ and $\Gamma=115\pm 20^{36}_{-32}$ MeV. The reported
mass and decay width changes a little in the published
paper~\cite{belle3}. $X(2690)$ was also reported by
BaBar~\cite{babar3}, but the significance of the signal was not
stated. This state is included in PDG~\cite{pdg08} with $M=2690\pm
7$ MeV, $J^P=1^-$ and full width $\Gamma=110\pm 27$ MeV.

$D_{sJ}(2860)$ was firstly reported by BaBar~\cite{babar3} in
$D_{sJ}(2860)\to D^0K^+~,~D^+K^0_s$ with $M=2856.6\pm 1.5(stat)\pm
5.0(syst)$ and $\Gamma=48\pm 7(stat)\pm 10(syst)$ MeV. It has
natural spin-parity: $J^P=0^+,~1^-,~\cdots$.

Before we proceed with the analysis of these two states, a brief
theoretical introduction to the features of the $D_s$ mesons is
suitable. When we have a look at previous theoretical computations,
we notice that two kinds of classification schemes about the $D_s$
are popularly employed: the $D_s$ is classified by
$^{2S+1}L_J$~\cite{gi,gk,zhao} or by $^jL_J$(where the HQET was
incorporated in and $j$ is the total momentum of the light degree of
freedom)~\cite{rq,pe,other1,other2}. That is to say, two different
bases of the wave functions of $D_s$ are employed. The relations
between these two bases are determined by the Clebsch-Gordan
coefficients~\cite{cj}. In the D-wave multiplets of $D_s$, the
$^{3\over 2}D_1$ and the $^{5\over 2}D_3$ in the $^jL_J$ basis
correspond to the $^3D_1$ and the $^3D_3$ in the $^{2S+1}L_J$ basis,
respectively. For the $D_2$ states,
\begin{eqnarray*}
^{3\over 2}D_2=\sqrt{3\over 5}~^3D_2-\sqrt{2\over 5}~^1D_2,\\
^{5\over 2}D_2=\sqrt{2\over 5}~^3D_2+\sqrt{3\over 5}~^1D_2.
\end{eqnarray*}

In experiments, some states are the pure $^{2S+1}L_J$ states(or the
pure $^jL_J$ states), some states are the mixing states of the
$^{2S+1}L_J$ states(or the mixing states of the $^jL_J$ states). For
example, the lowest observed S-wave $0^-$ $D_s$ must be the
$^1S_0$(or the $^{1\over 2}S_0$), while the lowest observed D-wave
$J^P=2^-$ $D_s$ should be the mixing of the $^3D_2$ and the
$^1D_2$(or the mixing of the $^{3\over 2}D_2$ and the $^{5\over
2}D_2$).

The features of the pure states may be simple, but the features of
the mixing states are much more complicated. Only when the exact
components of the mixing states are clear, we can understand the
features of these states. Unfortunately, the exact components of
some observed states are not well understood. Therefore, people is
usually careful with his conclusion when he observes a new state,
which may requires much more explorations. In this Letter, the decay
features of $D_s$ are touched and the spectrum is concentrated on.

The D-wave multiplets of $D_s$ and the radially excited $D_s$ have
been studied for a long time. Though the predictions of the masses
or the decay features are sometimes different in different models,
these theoretical explorations provide much information about the
quark dynamics.

The $1^3D_1$ is predicted to have mass $\approx 2900$
MeV~\cite{gi,pe} and total decay width $\Gamma=331$ MeV~\cite{cs}.
The transition $D_s(1~^3D_1)\to DK$ is predicted to have decay width
$26.1$ MeV in Ref.~\cite{pe}, and broader width $120$ MeV in
Ref.~\cite{cs}.

The $1^3D_3$ is expected to have mass $\approx 2920$
MeV~\cite{gi,pe} and total width $\Gamma=222$ MeV~\cite{cs}. The
predicted decay width of $D_s(1~^3D_3)\to DK$ in Ref.~\cite{pe} is
different with that in Ref.~\cite{cs}. It is $11.4$ MeV in
Ref.~\cite{pe} and $82$ MeV in Ref.~\cite{cs}.

The lowest three radially excited states are the $2~^1S_0$, the
$2~^3S_1$ and the $2~^3P_0$. The $2~^3S_1$ is predicted to have mass
$\approx 2700$ MeV~\cite{gi,rq} and total width $\Gamma=105$
MeV~\cite{cs}. The predicted decay width of $D_s(2~^3S_1)\to DK$ in
Ref.~\cite{pe}($21.1$ MeV) is similar to that in Ref.~\cite{cs}($17$
MeV).

The $2~^3P_0$ is predicted to have mass $\approx 3067$ MeV~\cite{pe}
and total width $\Gamma=90$ MeV~\cite{ctls}. The decay
$D_s(2~^3P_0)\to DK$ is predicted to have width $74.1$ MeV~\cite{pe}
or $80$ MeV~\cite{ctls}.

Based on these theoretical explorations and further analyses, some
assignments of $D_{s1}(2700)^\pm$ and $D_{sJ}(2860)$ have been
suggested. In existing literatures, $c\bar s$ assignments are most
advocated. These assignments will be examined in a phenomenological
way. Especially, some implications relevant to these assignments are
explored.

\section{Possible assignments of $D_{s1}(2700)^\pm$ and
$D_{sJ}(2860)$}

Comparing with the theoretical predictions of the masses and the
decay features, it is natural to explain $D_{s1}(2700)^\pm$ and
$D_{sJ}(2860)$ as the D-wave $D_s$ or the radially excited $D_s$.

In Ref.~\cite{zhu}, the authors believe that $D_{s1}(2700)^\pm$ is
probable the $1^-(1~^3D_1)$ $D_s$ through the study of its strong
decay. If this assignment is true, the mass of $D_{s1}(2700)^\pm$ is
$\approx 200$ MeV lower than theoretical predictions~\cite{gi,pe}.
In Refs.~\cite{cfn,zhu}, $D_{sJ}(2860)$ is interpreted as the
$3^-(1~^3D_3)$ $D_s$ .

As the candidates of the radial excitations, $D_{s1}(2700)^\pm$ is
interpreted as the $1^-(2~^3S_1)$ $D_s$~\cite{ctls}(first radial
excitation of the $D^\star_s(2112)^\pm$), and $D_{sJ}(2860)$ is
interpreted as the $0^+(2~^3P_0)$ $D_s$~\cite{br,zhu}(first radial
excitation of the $D^{\star}_{s0}(2317)^\pm$). If $D_{sJ}(2860)$ is
the $0^+(2~^3P_0)$ $D_s$, its mass is also $\approx 200$ MeV lower
than theoretical prediction~\cite{pe}.

Theoretical predictions of the D-wave and the radially excited S,
P-wave $D_s$ are summarized in Table.~\ref{table_1}. To compare
theoretical predictions explicitly, we show the results obtained in
Refs.~\cite{gi,rq,pe,other5}. Both the notation
$n^{2S+1}L_J$~\cite{gi,other5} and the notation $j^P$~\cite{rq,pe}
are employed for the classification of $c\bar s$. Accordingly, all
the quantum numbers $J^P$, $j^P$ and $n^{2S+1}L_J$ are listed. A
parenthesis is put on the $j^P$ for the $1D_2$ and the $2P_1$ states
to emphasize that there is no one-to-one correspondence between the
$n^{2S+1}L_J$ and the $^jL_J$(or $j^P$) classification scheme. The
dash "-" indicates that the mass of the corresponding state has not
been predicted, and the "$?$" indicates that there is not observed
candidate corresponding to the assignment. Possible assignments of
$D_{s1}(2700)^\pm$ and $D_{sJ}(2860)$ are denoted.

\begin{table}
\begin{tabular}{llllllll}
Candidates & $J^P$ & $j^P$ &  $n^{2S+1}L_J$ & \cite{gi} & \cite{rq}
& \cite{pe} & \cite{other5}\\
\hline\hline $D_{s1}(2700)$ & $1^-$ & ${3\over 2}^-$ & $1^3D_1$ & 2.90 & - & 2.913 & 2.714\\
? & $2^-$ & $({5\over 2}^-)$ & $1^3D_2$ & - & - & 2.900 & 2.789\\
? & $2^-$ & $({3\over 2}^-)$ & $1^1D_2$ & - & - & 2.953 & 2.827\\
$D_{sJ}(2860)$ & $3^-$ & ${5\over 2}^-$ & $1^3D_3$ & 2.92 & - & 2.925 & 2.903\\
\hline\hline ? & $0^-$ & ${1\over 2}^-$ & $2^1S_0$ & 2.67 & 2.670 & 2.700 & - \\
$D_{s1}(2700)$ & $1^-$ & ${1\over 2}^-$ & $2^3S_1$ & 2.73 & 2.716 & 2.806 & -\\
\hline\hline
$D_{sJ}(2860)$ & $0^+$ & ${1\over 2}^+$ & $2^3P_0$ & - & - & 3.067 & -\\
$?$ & $1^+$ & $({3\over 2}^+)$ & $2^3P_1$ & - & - & 3.114 & -\\
$?$ & $1^+$ & $({1\over 2}^+)$ & $2^1P_1$ & - & - & 3.165 & -\\
$?$ & $2^+$ & ${3\over 2}^+$ & $2^3P_2$ & - & - & 3.157 & -\\
\hline\hline
\end{tabular}
\caption{Spectrum of the $1D$, the $2S$ and the $2P$ $D_s$(GeV).}
\label{table-1}
\end{table}

These two states seem a little "exotic" to the assignments. On one
hand, some experimentally measured decay widths deviate largely from
theoretical predictions. On the other hand, the masses of
$D_{s1}(2700)^\pm$ and $D_{sJ}(2860)$ are $\approx 200$ MeV lower
than the predictions of the $1^-(1~^3D_1)$ and the $0^+(2~^3P_0)$
$D_s$.

These "exotic" features imply that existing models may not be valid,
or models are valid but need improvement, or the interpretations of
the states are incorrect. Therefore it is necessary to study the
"exotic" features.

The character of lower masses is well explained in the coupled
channels analysis~\cite{br}. This feature is also well explained in
a mass loaded flux tube model~\cite{sw,other5}. In the model, the
orbitally excited masses of $D$ and $D_s$ are computed through
\begin{eqnarray}
E=M+\sqrt{{\sigma L\over 2}}+2^{1\over 4}\kappa L^{-{1\over
4}}m^{3\over 2}+a \vec {L}\cdot \vec{S}.
\end{eqnarray}

The predicted masses for the $1^-(1~^3D_1)$($\approx 2714\pm 30$
MeV) and the $3^-(1~^3D_3)$ $D_s$($\approx 2903\pm 40$ MeV) agree
well with that of $D_{s1}(2700)$ and $D_{sJ}(2860)^\pm$ states,
respectively.

The masses of the $1~^3D_2$($\approx 2789\pm 44$ MeV) and the
$1~^1D_2$($\approx 2827\pm 44$ MeV) are predicted in
Ref.~\cite{other5}. Concern about the mixing between the the $^3D_2$
and the $^1D_2$, a $2^-$ $D_2$ charmed strange meson(denoted as
$D_{s2}(2800)$) with mass $\approx 2800$ MeV is expected.

The widths of some transitions of the $D_s(2^-~^{5\over 2}D_2)$
($\approx 2900$ MeV) and the $D_s(2^-~^{3\over 2}D_2)$ ($\approx
2953$ MeV) have been computed in Ref.~\cite{pe}. In terms of
Eq.~(33) in Ref.~\cite{pe}, the new results are obtained in
Table.~\ref{table-2}(with mass correction considered), where $l_x$
is the angular momentum of the final light mesons.

\begin{table}
\begin{tabular}{llll}
\hline\hline state & ~~~~~~~~channel & width(MeV) & $l_x$ \\
\hline\hline
$D_{s2}(2800)$ & $D_s(2^-~^{5\over 2}D_2)\to D^\star K$ & 11.0 & 3 \\
$D_{s2}(2800)$ & $D_s(2^-~^{5\over 2}D_2)\to D^\star_s \eta$ & 1.5 & 3 \\
\hline\hline
$D_{s2}(2800)$ & $D_s(2^-~^{3\over 2}D_2)\to D^\star K$ & 36.9 & 1 \\
$D_{s2}(2800)$ & $D_s(2^-~^{3\over 2}D_2)\to D^\star_s \eta$ & 17.3 & 1 \\
\hline\hline
\end{tabular}
\caption{Partial widths of some transitions of $D_{s2}(2800)$}
\label{table-2}
\end{table}

$D_{s2}(2800)$) is the mixing of the $^{3\over 2}D_2$ and the
$^{5\over 2}D_2$. When the mixing effect is taken account of, the
decay channel $D_{s2}(2800)\to D^\star K$ has width $\approx 11-37$
MeV and the channel $D_{s2}(2800)\to D^\star_s \eta$ has width
$\approx 2-17$ MeV.

Radial excitation of $D$ and $D_s$ has not yet been discovered, but
the mass of some radial excitations have been computed in models. In
what follows, we avoid the model dependent computation of the
spectrum and proceed with a phenomenological analysis.

As explored in Ref.~\cite{aas}, in the mass region up to $M<2400$
MeV, a mass equation about the radially excited mesons holds with a
good accuracy(trajectories on $(n,M^2)$-plots)
\begin{eqnarray}\label{traj}
M^2=M^2_0+(n-1)\mu^2
\end{eqnarray}
where $M_0$ is the mass of the basic meson, $M$ is the mass of the
radial excitation, $n$ is the radial quantum number, and $\mu^2$ is
the slope parameter of the trajectory(approximately the same for all
trajectories).

For the $D_s$, it is easy to observe in PDG~\cite{pdg08}
\begin{eqnarray}\label{gap}
M^2(D_{s1}(2700)^\pm)-M^2(D^\star_s(2112)^\pm)=2.78~GeV^2,\\
M^2(D_{sJ}(2860))-M^2(D^\star_{s0}(2317)^\pm)=2.79~GeV^2.
\end{eqnarray}

Obviously, the assignments of $D_{s1}(2700)^\pm$ and $D_{sJ}(2860)$
with the $1^-(2~^3S_1)$ and the $0^+(2~^3P_0)$ radially excited
$D_s$, respectively, are perfectly consistent with the
Eq.~(\ref{traj}).

Eq.~\ref{gap} indicates explicitly that Eq.~(\ref{traj}) is very
possible to hold in the mass region up to $D_s$ mesons. This
relation is supposed to hold and will be taken used of in our
analysis. Therefore, the masses of the $0^-$ and the $1^+$ radially
excited $D_s$ are easily obtained(see Table.~\ref{table-3}). The
$0^-$ and the $1^+$ $D_s$ are denoted as $D^\prime_s(2582)$ and
$D^\prime_{s1}(2973)$, respectively. $D^\prime_s(2582)$ is the
$0^-~2~^1S_0$(or the $2~{1\over 2}^-$). $D^\prime_{s1}(2973)$ is the
mixing of $1^+~2~^3P_1$ and $1^+~2~^1P_1$(or the mixing of the
$2~{1\over 2}^+$ and $2~{3\over 2}^+$).
\begin{table}
\begin{tabular}{lllll}
\hline\hline n & $0^-$ & $1^-$ & $0^+$ & $1^+$ \\
\hline\hline
1 & $D_s(1969)^\pm$ & $D^\star_s(2112)^\pm$ & $D^\star_{s0}(2317)^\pm$ & $D_{s1}(2460)^\pm$ \\
2 & $D^\prime_s(2582)$ & $D_{s1}(2700)^\pm$ & $D_{sJ}(2860)$ & $D^\prime_{s1}(2973)$\\
\hline\hline
\end{tabular}
\caption{Ground and first radially excited $(0^-,1^-)$ and
$(0^+,1^+)$ $D_s$} \label{table-3}
\end{table}

It is interesting to notice that the chiral doubling relation still
exist in radially excited $D_s$. They have splitting $\approx 280$
MeV
\begin{eqnarray*}
D^\prime_{s1}(2973) -D_{s1}(2700)^\pm \approx D_{sJ}(2860)
-D^\prime_s(2582).
\end{eqnarray*}
The splitting is smaller than the splitting($\approx 380$ MeV) of
the ground states
\begin{eqnarray*}
D_{s1}(2460)^\pm -D^\star_s(2112)^\pm \approx D^\star_{s0}(2317)^\pm
-D_s(1969)^\pm.
\end{eqnarray*}

The widths of some decay channels of the $D_s(1^+~^{3\over 2}P_1)$
($\approx 3114$ MeV) and the $D_s(1^+~^{1\over 2}P_1)$ ($\approx
3165$ MeV) have been computed in Ref.~\cite{pe}.

\begin{table}
\begin{tabular}{llll}
\hline\hline state & ~~~~~~~channel & width(MeV) & $l_x$ \\
\hline\hline
$D^\prime_{s1}(2973)$ & $D_s(1^+~^{3\over 2}P_1)\to D^\star K$ & 36.6 & 2 \\
$D^\prime_{s1}(2973)$ & $D_s(1^+~^{3\over 2}P_1)\to D^\star_0 K$ & 5.5 & 1 \\
$D^\prime_{s1}(2973)$ & $D_s(1^+~^{3\over 2}P_1)\to D^\star_s \eta$ & 11.1 & 2 \\
$D^\prime_{s1}(2973)$ & $D_s(1^+~^{3\over 2}P_1)\to D^\star_1 K$ & 4.5 & 1 \\
\hline\hline
$D^\prime_{s1}(2973)$ & $D_s(1^+~^{1\over 2}P_1)\to D^\star K$ & 72.1 & 0 \\
$D^\prime_{s1}(2973)$ & $D_s(1^+~^{1\over 2}P_1)\to D^\star_0 K$ & 42.6 & 1 \\
$D^\prime_{s1}(2973)$ & $D_s(1^+~^{1\over 2}P_1)\to D^\star_2 K$ & 41.3 & 1 \\
$D^\prime_{s1}(2973)$ & $D_s(1^+~^{1\over 2}P_1)\to D^\star_s \eta$ & 45.1 & 0 \\
\hline\hline
\end{tabular}
\caption{Partial widths of some transitions of
$D^\prime_{s1}(2973)$}\label{table-4}
\end{table}

The decay widths of $D^\prime_s(2582)$ could be obtained directly
from Ref.~\cite{pe}. It is a little complicated to compute the decay
widths of $D^\prime_{s1}(2973)$. Only when the transition widths of
the $^{1\over 2}P_1$ and the $^{3\over 2}P_1$ have been
computed(Table. ~\ref{table-4}), can the transition widths of
$D^\prime_{s1}(2973)$ be obtained from their mixing.

The total width of $D^\prime_{s1}(2973)$ is broad. For separate
channel, the channel $D^\prime_{s1}(2973)\to D^\star K$ has width
$\approx 37-72$ MeV, the channel $D^\prime_{s1}(2973)\to D^\star_0
K$ has width $\approx 6-43$ MeV and the channel
$D^\prime_{s1}(2973)\to D^\star_s \eta$ has width $\approx 5-45$
MeV.

Of course, once $D_{s1}(2700)^\pm$ is pinned down as the
$1^-(2^3S_1)$, the exotic $D_{sJ}(2632)^+$ reported by
SELEX~\cite{selex} becomes a supernumerary one. In
Ref.~\cite{zhang}, the exotic feature of the mass of
$D_{sJ}(2632)^+$ has been mentioned in a similar way. How to put
this state into the $D_s$ zoo deserves more exploration.

\section{Discussions and conclusions}

Some features of $D_{s1}(2700)^\pm$ and $D_{sJ}(2860)$ are really
different with previous predictions. In our analysis, these features
could be explained in the conventional quark model. "Exotic"
interpretation outside of the $c\bar s$ assignments may not be
necessary.

$D_{s1}(2700)^\pm$ and $D_{sJ}(2860)$ may be the orbitally excited
D-wave $1^-~(j^P={3\over 2}^-)$ and $3^-~(j^P={5\over 2}^-)$,
respectively. In these assignments, another $D_{s2}(2800)$ with mass
$\approx 2800$ MeV is expected. It is the mixing of the $^3D_2$ and
the $^1D_1$(or the mixing of the $^{3\over 2}D_2$ and the $^{5\over
2}D_2$). The decay $D_{s2}(2800)\to D^\star K$ has width $\approx
11-37$ MeV and the channel $D_{s2}(2800)\to D^\star_s \eta$ has
width $\approx 2-17$ MeV.

$D_{s1}(2700)^\pm$ and $D_{sJ}(2860)$ are very possible the first
radially excited $1^-~(j^P={1\over 2}^-)$ and $0^+~(j^P={1\over
2}^+)$, respectively. In this case, other two radially excited
$0^-~D^\prime_s(2582)$ and $1^+~D^\prime_{s1}(2973)$ are expected.
These two states are the radial excitations of $D_s(1969)^\pm$ and
$D_{s1}(2460)^\pm$. The chiral doubling relation may exist in
radially excited $D_s$, the splitting between the parity
partners(the $(0^-,1^-)$ and the $(0^+,1^+)$) is $\approx 280$ MeV.
$D^\prime_{s1}(2973)$ has a broad total width.

The component of an observed mixed state is usually not clear, it is
an important origin to the "exotic" properties of the observed
state. How to explain an observed state in terms of the features of
the pure states deserves more exploration.

Acknowledgment: This work is supported by the National Natural
Science Foundation of China under the grant: 10775093.

\end{document}